\documentclass[12pt]{article}
\usepackage{latexsym}
\usepackage{amsmath}
\usepackage{bm}
\usepackage[dvipdfmx]{graphicx}
\usepackage{color}
\usepackage{amssymb}
\usepackage{cite}
\usepackage{amsfonts, amscd, amsthm}
\usepackage[all]{xy}
\usepackage{here}
\usepackage{ulem}

\renewcommand{\theequation}{\thesection.\arabic{equation}}

\makeatletter
\@addtoreset{equation}{section}
\renewcommand{\theequation}{\thesection.\@arabic\c@equation}
\makeatother

\makeatletter
\renewcommand\appendix{\par
  \setcounter{section}{0}%
  \setcounter{subsection}{0}%
  \gdef\thesection{Appendix \@Alph\c@section }
  \renewcommand{\theequation}
  {\Alph{section}.\arabic{equation}}
}
\makeatother

\makeatletter
\newcounter{subeqncnt}
\def\thesubeqncnt{\alph{subeqncnt}}
\def\subequations{\begingroup%
\stepcounter{equation}\edef\@tempa{\theequation}%
\let\c@equation\c@subeqncnt\c@subeqncnt\z@
\edef\theequation{\@tempa\noexpand\thesubeqncnt}}

\makeatother

\begin{document}

\titlepage

\title{Elliptic Solutions for Higher Order KdV Equations} 
\author{Masahito Hayashi\thanks{masahito.hayashi@oit.ac.jp}\\
Osaka Institute of Technology, Osaka 535-8585, Japan\\
Kazuyasu Shigemoto\thanks{shigemot@tezukayama-u.ac.jp} \\
Tezukayama University, Nara 631-8501, Japan\\
Takuya Tsukioka\thanks{tsukioka@bukkyo-u.ac.jp}\\
Bukkyo University, Kyoto 603-8301, Japan\\
}
\date{\empty}


\maketitle
\abstract{
We study higher order KdV equations from 
the GL(2,$\mathbb{R}$) $\cong$ SO(2,1) Lie group point of view.
We find elliptic solutions of higher order KdV equations  
up to the ninth order. 
We argue that the main structure of the 
trigonometric/hyperbolic/elliptic $N$-soliton solutions 
for higher order KdV equations is the same 
as that of the original KdV equation. Pointing out that the 
difference is only the 
time dependence, we find $N$-soliton solutions of higher order
KdV equations can be constructed from those 
of the original KdV equation by properly replacing the time-dependence.
We discuss that there always exist elliptic solutions 
for all higher order KdV equations. 
}

\section{Introduction} 
\setcounter{equation}{0}

The soliton system is taken an interest in for a long time   
by considering that the soliton equation is the concrete example of the 
exactly solvable nonlinear differential 
equation~\cite{Gardner,Lax,Zakhrov,Ablowitz,Wahlquist,Wadati1,
Wadati2,Hirota1,Hirota2,Sato,Date1,Weiss}.
Nonlinear differential equation relates to the interesting 
non-perturbative phenomena, so that studies of the soliton system
are important to unveil mechanisms of various interesting physical 
phenomena such as those in superstring theories.
It is quite surprising that such nonlinear soliton equations 
can be exactly solvable and have $N$-soliton solutions.
Then we have a dogma that there must be the Lie group structure 
behind the soliton system, which is a key stone to make nonlinear 
differential equations exactly solvable.

For the KdV soliton system, the Lie group structure is implicitly 
built in the Lax operator $L=\partial_x^2-u(x,t)$. 
In order to see the Lie group structure, it is appropriate to formulate by using
the linear differential operator $\partial_x$ as the Schr\"{o}dinger 
representation of the Lie algebra, which naturally comes to use the 
AKNS formalism~\cite{Ablowitz} for the Lax equation
\begin{align}
\frac{\partial}{\partial x}   \left(\begin{array}{c} \psi_1(x, t)    \\  \psi_2(x, t)
  \end{array}\right)  &=  \left(\begin{array}{cc}
  a/2 & -u(x, t) \\  -1 & -a/2  \end{array}\right)
  \left(\begin{array}{c}  \psi_1(x, t)    \\   \psi_2(x, t) \end{array}\right)  .
\nonumber
\end{align} 
Then the Lie group becomes GL(2,$\mathbb{R}$) $\cong$ SO(2,1) for 
the KdV equation. 
An addition formula for elements of this Lie group is the well-known 
KdV type B\"{a}cklund transformation.

In our previous papers~\cite{Hayashi1,Hayashi2,Hayashi3,Hayashi4},  we have studied 
 GL(2,$\mathbb{R}$) $\cong$ SO(2,1) Lie group approach for the unified soliton 
systems of KdV/mKdV/sinh-Gordon equations. 
Using the well-know KdV type B\"{a}cklund transformation as the addition 
formula, we have algebraically constructed $N$-soliton solutions
from various trigonometric/hyperbolic 1-soliton 
solutions~\cite{Hayashi1,Hayashi3,Hayashi4}. 
Since the Lie group structure of KdV equation is the 
GL(2,$\mathbb{R}$) $\cong$ SO(2,1), which has elliptic solution, 
we expect that 
elliptic $N$-soliton solutions for the KdV equation 
can be constructed 
by using the B\"{a}cklund transformation as the addition formula.  
We then really have succeeded in constructing elliptic $N$-soliton 
solutions~\cite{Hayashi2}. 

We can interpret this fact in the following way: 
The KdV equation, which
is a typical 2-dimensional soliton equation, has the SO(2,1) Lie group 
structure and the well-known KdV type B\"{a}cklund transformation 
can be interpreted as the addition formula of this Lie group. Then the 
elliptic function appears as a representation of the B\"{a}cklund 
transformation.
While, 2-dimensional Ising model, which is a typical 2-dimensional
statistical integrable model, has the SO(3) Lie group structure and the 
Yang-Baxter relation can be interpreted as the addition formula of this 
Lie group. Then the elliptic function appears as a representation of the
Yang-Baxter relation, which is equivalent to the addition formula of the 
spherical trigonometry~\cite{Shigemoto1,Shigemoto2}.
In 2-dimensional integrable, soliton, and statistical models, 
there is the SO(2,1)/SO(3) Lie group structure behind the model.  
As representations of the addition formula, the B\"{a}cklund 
transformation, and the Yang-Baxter relation, there appears an algebraic
function such as the trigonometric/hyperbolic/elliptic functions,
which is the key stone to make the 2-dimensional integrable model
into the exactly solvable model. 

In this paper, we consider Lax type higher order KdV equations 
and study trigonometric/hyperbolic/elliptic solutions.
So far special hyperelliptic solutions for more than  
the fifth order KdV equation have been vigorously studied by formulating 
it into the Jacobi's inversion 
problem~\cite{Burchnall,Date2,McKean,Its,Dubrobin,Krichever}.
Since the Lie group structure  GL(2,$\mathbb{R}$) $\cong$ SO(2,1) 
and the B\"{a}cklund transformation are common even for  
higher order KdV equations, we expect that there always exist  
elliptic solutions even for higher order. 
Then we study to find elliptic solutions up to the ninth order KdV
equation, 
instead of special hyperelliptic solutions. 
We would like to conclude that we always have 
elliptic solutions for all higher order KdV equations.

As the application of the third order KdV equation, this equation 
is first obtained 
in the analysis of shallow water solitary wave~\cite{KdV}.
Even recently, the third order KdV equation becomes 
important in the analysis of various non-linear phenomena.
For example, in the recent interesting works,  
the third order KdV equation comes out in the analysis 
of the non-linear acoustic solitary wave in the 
electron-ion plasma~\cite{Saha1,Saha2,Saha3,Saha4}.
As the application of the higher order KdV equation, 
some special fifth order KdV equation(KdV5), which is 
different from the Lax type equation,  is 
recently experimentally and theoretically  interested in.
This KdV5 equation comes out in the analysis of 
various non-linear phenomena, such as 
cold collisionless plasma~\cite{Kakutani}, gravity-capillary 
wave~\cite{Hunter}, shallow water wave with surface 
tension~\cite{Sun} etc.
Theoretically, it is shown that Camassa-Holm equation is 
transformed into this KdV5 equation~\cite{Holm1,Holm2} 
and multi-soliton solutions is obtained~\cite{Wazwaz1}. 
In this way, the KdV equation becomes important
in the analysis of various non-linear phenomena.

The paper is organized as follows: 
In section $2$, we study trigonometric/hyperbolic solutions for   
higher order KdV equations.
We construct elliptic solutions for  
higher order KdV equations in section $3$.
In section $4$, we consider the KdV type B\"{a}cklund transformation
as an addition formula for solutions of the Weierstrass type elliptic 
differential equation. 
In section $5$, we study special 1-variable hyperelliptic 
solutions, and we discuss 
a relation between such special 1-variable hyperelliptic solutions 
and our elliptic solutions. 
We devote a final section to summarize this paper and to give discussions.

\section{Trigonometric/hyperbolic solutions for the Lax type 
higher order KdV equations} 
\setcounter{equation}{0}

Lax pair equations for higher order KdV equations are given by 
\begin{eqnarray}
&& L\psi=\frac{a^2}{4} \psi ,
\label{2e1}\\
&& \frac{\partial \psi}{\partial t_{2n+1}} =B_{2n+1} \psi , 
\label{2e2}
\end{eqnarray}
where $L=\partial_x^2 -u$.
By using the pseudo-differential 
operator $\partial_x^{-1}$,  $B_{2n+1}$ are constructed from $L$
in the form~\cite{Gelfand,Deckey} 
\begin{equation}
B_{2n+1}=\left(\mathcal{L}^{2n+1}\right)_{\geq 0}=\partial_x^{2n+1}
-\frac{2n+1}{2} u \partial_x^{2n-1} +\cdots ,
\label{2e3}
\end{equation}
with 
$$
\mathcal{L}=L^{1/2}=\partial_x-\frac{u}{2} \partial_x^{-1} 
+\frac{u_x}{4} \partial_x^{-2}+\cdots ,
$$
where we denote ``$\geq0$'' to take positive differential operator parts or
function parts for general pseudo-differential operators. 
The integrability condition gives 
higher order KdV equations 
\begin{eqnarray}
\frac{\partial L}{\partial t_{2n+1}}=[B_{2n+1}, \, L] .
\label{2e5}
\end{eqnarray} 
As these higher order KdV equations comes from the Lax formalism, these 
higher order KdV equations are called the Lax type. There are various
higher order KdV equations such as the Sawada-Kotera 
type, which is the higher order generalization of the Hirota form  
KdV equation~\cite{Wazwaz}. 
As operators $B_{2n+1}$ 
are constructed from $L$, higher order KdV equations also have the same 
Lie group structure GL(2,$\mathbb{R}$) $\cong$ SO(2,1) 
as that of the original KdV(=third order KdV) equation.
Using $u=z_x$, the KdV type B\"{a}cklund transformation is given in the form
\begin{eqnarray}
z'_x+z_x=-\frac{a^2}{2}+\frac{(z'-z)^2}{2} , 
\label{2e6}
\end{eqnarray}
which comes from Eq.(\ref{2e1}) only, so that it is valid even for the 
higher order KdV equations. 
In the Lie group approach to the soliton system, if we find 1-soliton 
solutions, we can construct $N$-soliton solutions from various 
1-soliton solutions by the B\"{a}cklund transformation Eq.(\ref{2e6}) 
as an addition formula of the Lie group.

For 1-soliton solution of Eq.(\ref{2e5}), if $x$ and $t_{2n+1}$ 
come in the combination 
$X^{(2n+1)}=\alpha x +\beta t_{2n+1}^{\gamma} +\delta$, 
then if $\gamma \ne 1$, the right-hand side of  Eq.(\ref{2e5}) is a function 
of only $X$, while the left-hand side is a function of $X$ and
$t$. Therefore,  
$\gamma=1$ is necessary, that is, $X=\alpha x +\beta t_{2n+1}+\delta$.  
$N$-soliton solutions are constructed from various 1-soliton solutions 
by the B\"{a}cklund transformation. Then the main structure of 
$N$-soliton solutions, which are expressed with $X_i^{(2n+1)}, (i=1,2, \cdots,N)$, 
takes the same functional forms in higher order KdV equations 
and in the original KdV equation. 
The difference is only the time dependence of 
$X_i=\alpha_i x +\beta_i t_{2n+1}+\delta_i ,(i=1,2, \cdots,N)$, that is,  
coefficients $\beta_i$. 
This is valid  not only for the trigonometric/hyperbolic $N$-soliton solutions 
but also for elliptic $N$-soliton solutions. 

For the trigonometric/hyperbolic $N$-soliton solutions, we can easily determine the 
time dependence without knowing details of $B_{2n+1}$. 
For dimensional analysis, we have $[\partial_x]=M$, $[u]=M^2$ in the unit of 
mass dimension $M$. Further, we notice that $[B_{2n+1}, \, L] $  does not 
contain differential operators but it contains only functions.  
Then we have 
\begin{eqnarray}
&& \frac{\partial u }{\partial t_{2n+1} } =\partial_x^{2n+1} u 
+\mathcal{O}(u^2) .
\label{2e7}
\end{eqnarray}
As Eq.(\ref{2e7}) is the Lie group type differential equation, we take the
Lie algebraic limit.
Putting $u=\epsilon \hat{u}$ first, 
Eq.(\ref{2e7}) takes in the form 
\begin{eqnarray}
\epsilon \frac{\partial \hat{u} }{\partial t_{2n+1} } =\epsilon \partial_x^{2n+1} \hat{u} 
+\mathcal{O}(\epsilon^2 \hat{u}^2) ,
\label{2e8}
\end{eqnarray}
and afterwards we take the limit $\epsilon \rightarrow 0$, which gives   
\begin{eqnarray}
 \frac{\partial \hat{u} }{\partial t_{2n+1} } =\partial_x^{2n+1} \hat{u} .
\label{2e9}
\end{eqnarray}
Then for trigonometric/hyperbolic solutions, we see that $x$ 
and $t_{2n+1}$ come in a combination
$X_i=a_i x+\delta_i$ $\rightarrow$ $X_i=a_i x + a_i^{2n+1}t_{2n+1}+\delta_i$
for 1-soliton solutions. 
In this way, the time-dependence for trigonometric/hyperbolic solutions
is easily determined without knowing details of $B_{2n+1}$.  
We can then obtain trigonometric/hyperbolic $N$-soliton solutions for the 
$(2n+1)$-th order KdV equation from the original KdV $N$-soliton solutions 
just by replacing  $X^{(3)}_i=a_i x + a^{3}_i t_{3}+\delta_i   \rightarrow
X^{(2n+1)}_i=a_i x + a^{2n+1}_i t_{2n+1}+\delta_i, (i=1,2, \cdots,N) $. 

For example, the original third order KdV equation is given 
by~\footnote{We use the notation 
$u_x=\partial_x u$, $u_{2x}=\partial_x^2 u$, $\cdots$, 
throughout the paper.} 
\begin{eqnarray}
u_{t_{3}}=u_{3x}-6 u u_x ,
\label{2e10} 
\end{eqnarray}
and the fifth order KdV equation is given by~\cite{Wazwaz},  
\begin{eqnarray}
u_{t_{5}}=u_{5x}-10u u_{3x} -20 u_x u_{2x}+30u^2 u_x .
\label{2e11} 
\end{eqnarray}
These two equations look quite different, but the 1-soliton solution for 
the third order KdV equation is given by 
$z=-a \tanh((ax+a^3 t+\delta)/2)$, while 1-soliton solution for 
the fifth order KdV equation is given by $z=-a \tanh((ax+a^5 t+\delta)/2)$.
In this way, even for any $N$-soliton solutions, we can obtain 
the fifth order KdV solution
from third order KdV solution just by replacing 
$X^{(3)}_i=a_ix+a_i^3 t+\delta_i \rightarrow X^{(5)}_i=a_ix+a_i^5 t+\delta_i$.
See more details in the Wazwaz's nice textbook~\cite{Wazwaz}.

However, as we explain in the next section, the way to determine 
the time dependence by taking the 
Lie algebraic limit does not applicable for elliptic solutions.

\section{Elliptic solutions for the Lax type higher order KdV equations} 
\setcounter{equation}{0}

We consider here elliptic 1-soliton solutions for higher order KdV equations
up to ninth order. 
We first study whether higher order KdV equations reduces to 
differential equations of the elliptic curves. 
If a differential equation of the elliptic curve exists, 
via dimensional analysis, 
$[\partial_x]=M$, $[u]=M^2$, $[k_3]=M^0$, $[k_2] =M^2$, 
$[k_1]=M^4$, and $[k_0]=M^6$, 
that must be the differential equation of the Weierstrass type 
elliptic curve
\begin{equation} 
{u_x}^2=k_3 u^3+k_2 u^2 +k_1 u +k_0 ,
\label{3e1}
\end{equation}
where $k_i (i=0, 1, 2, 3)$ are constants.
We cannot use the method to take the Lie algebraic limit to find 
the time dependence of the elliptic 1-soliton solution, because 
we cannot take $u \rightarrow 0$ as $k_0 \ne 0$ is essential 
in the elliptic case. 
By differentiating Eq.(\ref{3e1}), we have the following relations; 
\begin{subequations} 
\begin{align} 
u_{2x}&=\frac{3}{2} k_3 u^2+k_2 u+\frac{1}{2} k_1,
\label{3e2}\\ 
u_{3x}&=3 k_3 u u_x+k_2 u_x ,
\label{3e3}\\
u_{4x}&=3 k_3 u u_{2x}+3 k_3  {u_{x}}^2+k_2 u_{2x} ,
\label{3e4}\\
u_{5x}&=9 k_3 u_x u_{2x} +3 k_3 u u_{3x} +k_2  u_{3x} ,
\label{3e5}\\
u_{6x}&=12 k_3 u_x u_{3x} +9 k_3 {u_{2x}}^2 +3 k_3 u  u_{4x} +k_2  u_{4x} ,
\label{3e6}\\
u_{7x}&=30 k_3 u_{2x} u_{3x} +15 k_3 u_{x} u_{4x} 
+3 k_3 u u_{5x}+k_2 u_{5x} ,
\label{3e7}\\
u_{8x}&=45 k_3 u_{2x} u_{4x} +30 k_3 {u_{3x}}^2 +18 k_3 u_{x} u_{5x} 
+3 k_3  u u_{6x} +k_2 u_{6x} .
\label{3e8}
\end{align}
\end{subequations}
 
\vspace*{-5mm}

\subsection{Elliptic solution for the third order KdV(original KdV) equation}

The third order KdV (original KdV) equation is given by
\begin{eqnarray}
u_{t_{3}}=u_{3x}-6 u u_x=\left(u_{2x}-3u^2\right)_x  .
\label{3e9} 
\end{eqnarray}
We consider the 1-soliton solution, where $x$ and $t$ come in the 
combination $X=x+c_3 t_{3}+\delta$, then we have 
\begin{eqnarray}
u_{2x} -3 u^2-c_3 u=\frac{k_1}{2} ,
\label{3e10} 
\end{eqnarray}
where $k_1/2$ is an integration constant. Further multiplying $u_x$ and integrating, 
we have the following differential equation of the Weierstrass type elliptic curve  
\begin{eqnarray}
{u_x}^2=2 u^3+k_2 u^2+k_1 u +k_0, 
\label{3e11} 
\end{eqnarray}
where $k_2$, $k_1$, and $k_0$ are constants and $c_3$ is determined 
as $c_3=k_2$, which gives the time-dependence of the 1-soliton solution.
If we put $\wp=u/2+k_2/12$, we have 
the standard differential equation of the Weierstrass $\wp$ function type
\begin{equation}
\wp_x^2=4 \wp^3 -g_2\wp-g_3 ,
\label{3e12}
\end{equation}
with 
\begin{subequations} 
\begin{align}
g_2&={k_2}^2/12-k_1/2  ,
\label{3e13}
\\
g_3&=-{k_2}^3/216+k_1 k_2/24-k_0/4  .
\label{3e14}
\end{align}
\end{subequations}
Elliptic 1-soliton solution is given by 
\begin{equation}
u(x,t_3)=u(X^{(3)})=2 \wp(X^{(3)})-\frac{k_2}{6}  ,
\label{3e15}
\end{equation}
with 
$$
X^{(3)}=x+c_3 t_3+\delta, \quad c_3=k_2  .
$$
We sketch the graphs of the third order KdV solution in 
Figure 1. 
\begin{figure}[h!]
\qquad\quad
 \begin{minipage}{0.8\hsize}
  \begin{center}
\hspace{10mm}
   \includegraphics[width=100mm]{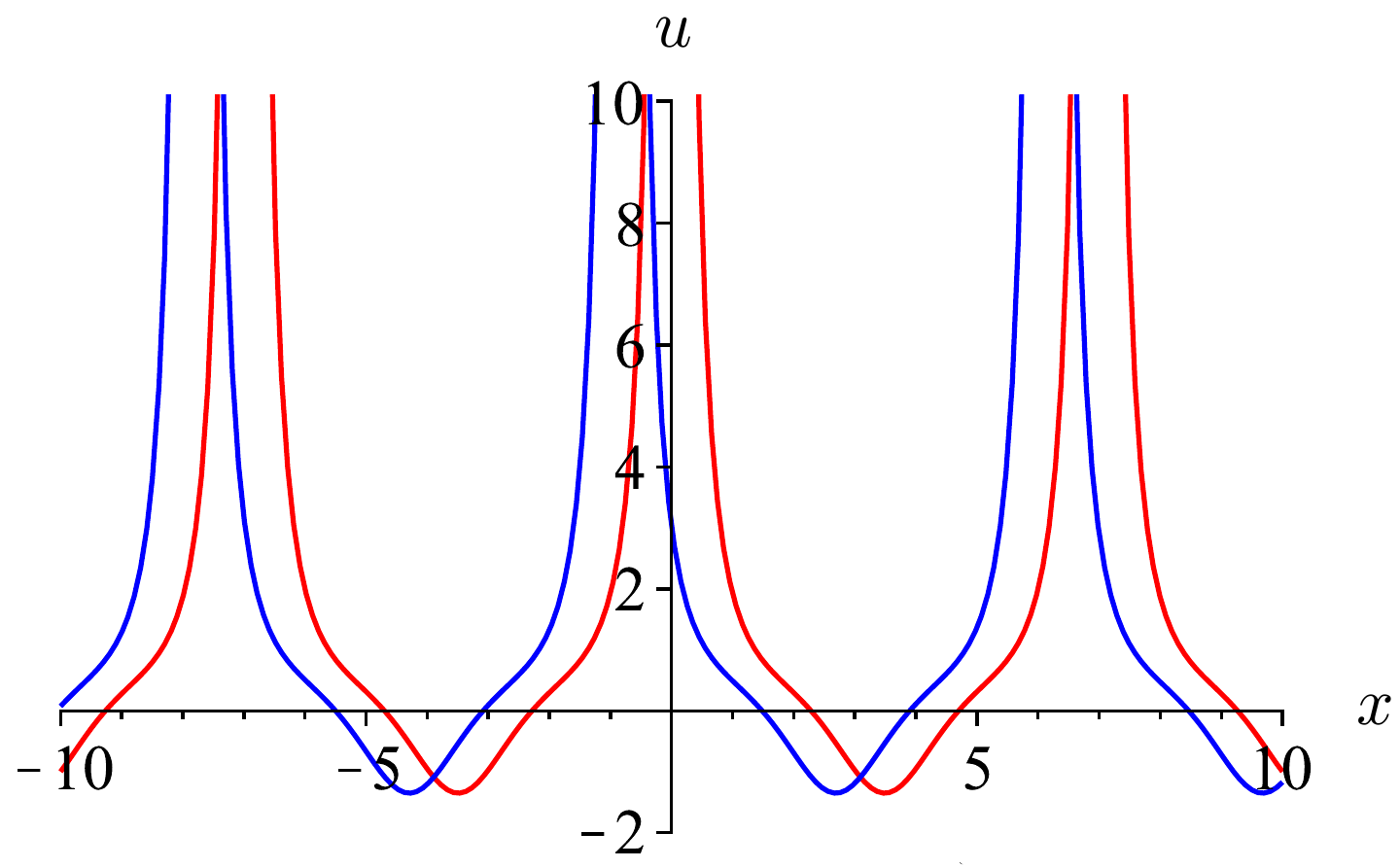}
  \end{center}
 \caption{The third order KdV solution.
The red line shows $u(x,0)$ and the blue line shows
 $u(x,1)$ with $k_0=1.2$, $k_1=-1.6$, $k_2=0.8$,
$\delta=0$. We can see the time-dependence.
}
  \label{fig1}
 \end{minipage}
\end{figure}
It should be noted that we must parametrize the differential 
equation of the Weierstrass type elliptic
curve by $k_2$, $k_1$, and $k_0$ instead of $g_2$ and $g_3$, 
because coefficients 
$c_{2n+1}$ in higher order KdV equations, which determine the time 
dependence, are expressed with $k_2$, $k_1$, and $k_0$.
According to the method of our previous paper, 
if we find various 1-soliton solutions, we can 
construct $N$-soliton solutions~\cite{Hayashi2}.

\subsection{Elliptic solution for the fifth order KdV equation}

The fifth order KdV equation is given by~\cite{Wazwaz},
\begin{eqnarray}
u_{t_{5}}-\left(u_{4x}-10u u_{2x} -5{u_{x}}^2+10u^3\right)_x=0  .
\label{3e17} 
\end{eqnarray}
We consider the elliptic solution, where $x$ and $t_{5}$ come 
in the combination of $X=x+c_5 t_5+\delta$, which gives
\begin{eqnarray}
c_{5}u-\left(u_{4x}-10u u_{2x} -5{u_{x}}^2+10u^3\right)+C=0,
\label{3e18} 
\end{eqnarray}
where $C$ is an integration constant. We will show that the above equation reduces to
the same differential equation of the Weierstrass type 
elliptic curve Eq.(\ref{3e1}).
Substituting Eq.(\ref{3e1}), $\cdots$, and Eq.(\ref{3e4}) into 
Eq.(\ref{3e18}) and comparing coefficients 
of $u^3$, $u^2$, $u^1$, and $u^0$, we have 4 conditions 
for 6 constants $k_3$, $k_2$, $k_1$, $k_0$, $c_5$, and $C$ in the form
\begin{subequations} 
\begin{align}
\text{i)}\ &\quad (k_3-2)(3k_3-2)=0,
\label{3e19}\\ 
\text{ii)}\ &\quad k_2 (k_3-2)=0,
\label{3e20}\\
\text{iii)}\ &\quad c_5=(9 k_3/2 -10) k_1+{k_2}^2, 
\label{3e21}\\
\text{iv)}\ &\quad C=(3k_3-5) k_0 +k_1 k_2/2. 
\label{3e22}
\end{align}
\end{subequations}

\vspace*{-5mm}
\noindent
Then we have two solutions
\begin{align}
\text{I)}\ &\quad k_3=2, \quad k_2, k_1, k_0:\text{arbitrary}, \quad   
c_5=-k_1+{k_2}^2, \quad C=k_0+k_1 k_2/2,
\label{3e23}\\ 
\text{II)}\ &\quad k_3=\frac{2}{3}, \quad k_2=0, \quad k_1, k_0:\text{arbitrary}, \quad   
c_5=-7 k_1, \quad  C= -3 k_0. 
\label{3e24}
\end{align}
We here take the most general solution, i.e.\ I) case, which gives the same 
differential equation of the elliptic curve 
${u_x}^2=2u^3+k_2 u^2+{k_2}^2 u+k_0$ as that of 
the third order KdV equation Eq.(\ref{3e11}) and $c_5$ is 
determined as $c_5=-k_1+{k_2}^2$. 
Elliptic 1-soliton solution is given by 
\begin{equation}
u(x,t_5)=u(X^{(5)})=2 \wp(X^{(5)})-\frac{k_2}{6} ,
\label{3e25}
\end{equation}
with
$$
X^{(5)}=x+c_5 t_5+\delta, \quad c_5=-k_1+{k_2}^2  .
$$
We sketch the graphs of the fifth order KdV solution in 
Figure 2. The red line
shows $u(x,0)$ and the blue line shows $u(x,1)$. We take 
$k_0=1.2$, $k_1=-1.6$, $k_2=0.8$,
$\delta=0$ in the graph. From this graph, we can see difference of 
the time-dependence of the solution between the third order solution
 and the fifth order solution.
\begin{figure}[h!]
\qquad\quad
 \begin{minipage}{0.8\hsize}
  \begin{center}
\hspace{10mm}
   \includegraphics[width=100mm]{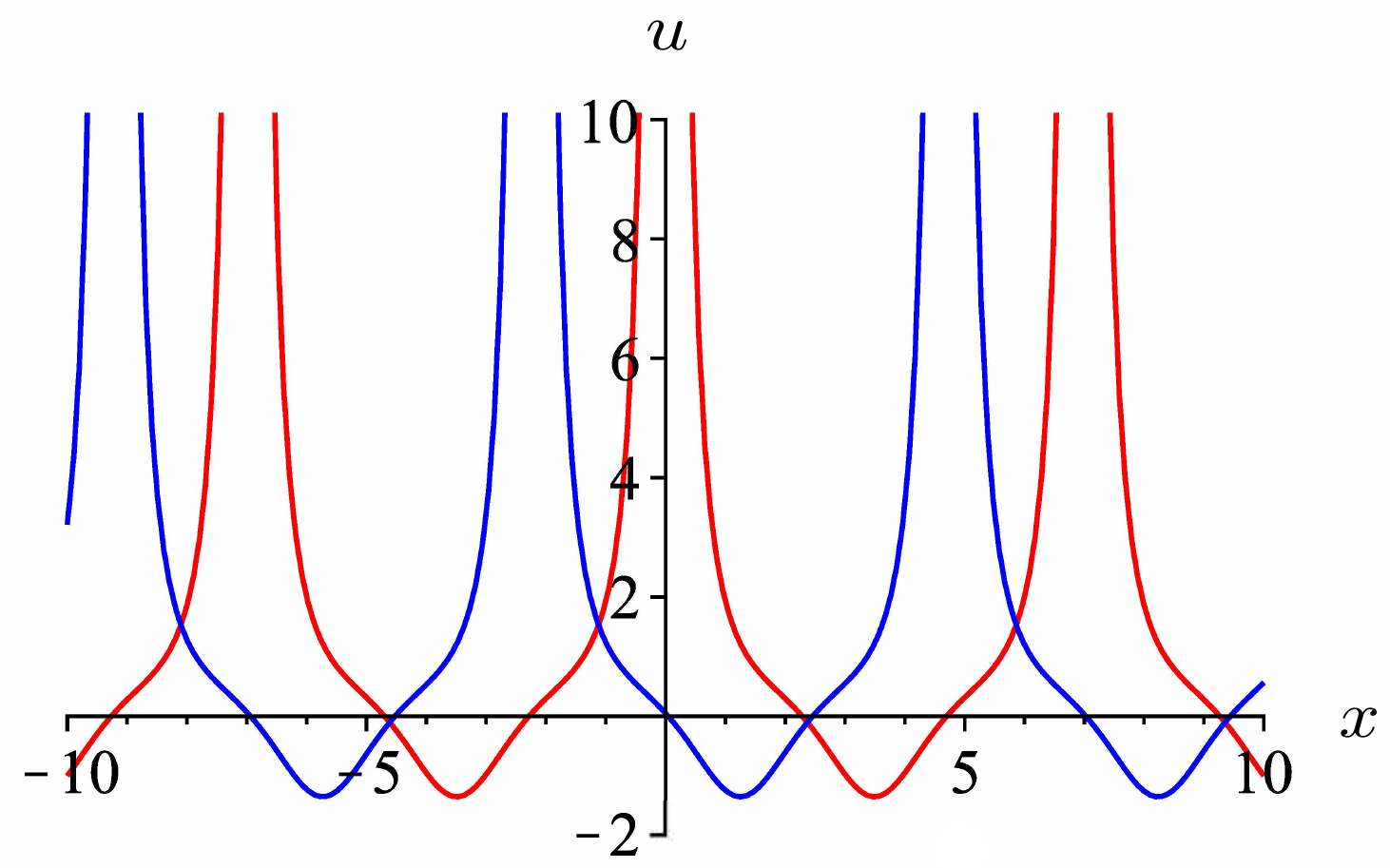}
  \end{center}
 \caption{The fifth order KdV solution.
The red line shows $u(x,0)$ and the blue line shows 
$u(x,1)$ with $k_0=1.2$, $k_1=-1.6$, $k_2=0.8$,
$\delta=0$. We can see the difference of the time-dependence 
between the third order solution and the fifth order solution.
}
  \label{fig2}
 \end{minipage}
\end{figure}

%
\subsection{Elliptic solution for the seventh order KdV equation}

The seventh order KdV equation is given by~\cite{Wazwaz}, 
\begin{equation} 
u_{t_{7}}-\left(u_{6x}-14u u_{4x} -28 u_x u_{3x}-21{u_{2x}}^2
+70 u^2 u_{2x} +70 u {u_x}^2 -35 u^4\right)_x=0  .
\label{3e27} 
\end{equation}
In this case, assuming that $x$ and $t_7$ come in the combination
of $X=x+c_7 t_7+\delta$, we have 
\begin{equation}
c_7u-\left(u_{6x}-14u u_{4x} -28 u_x u_{3x}-21{u_{2x}}^2
+70 u^2 u_{2x} +70 u {u_x}^2 -35 u^4\right)+C=0  .
\label{3e28} 
\end{equation}
Repeatedly substituting Eq.(\ref{3e1}),  $\cdots$, and Eq.(\ref{3e6}) 
into Eq.(\ref{3e28}) and comparing coefficients 
of $u^4$, $u^3$, $u^2$, $u^1$, and $u^0$, 
we have 5 conditions for 6 constants $k_3$, $k_2$, $k_1$, $k_0$, $c_7$, 
and $C$
of the form  
\begin{subequations} 
\begin{align}
\text{i)}\ &\quad (k_3-2)(3k_3-2)(3 k_3-1)=0,
\label{3e29}\\ 
\text{ii)}\ &\quad k_2(k_3-2)(3k_3-2)=0,
\label{3e30}\\
\text{iii)}\ &\quad k_1(k_3-2)(6k_3-5)+3 {k_2}^2(k_3-2)=0 ,
\label{3e31}\\
\text{iv)}\ &\quad c_7= (45 {k_3}^2-126 k_3 +70) k_0
 + (27k_3 -56) k_1 k_2 +{k_2}^3 ,
\label{3e32}\\
\text{v)}\ &\quad C=(15 k_3 -28)k_0 k_2
 + (9 k_3-21){k_1}^2/4 + k_1 {k_2}^2/2 .
\label{3e33}
\end{align}
\end{subequations}

\vspace*{-5mm}
\noindent
Then we get 3 solutions 
\begin{align}
\text{I)}\ &\quad k_3=2, \quad k_2, k_1, k_0:\text{arbitrary},  \quad
c_7=-2 k_0-2k_1 k_2+{k_2}^3, 
\nonumber\\
&\qquad 
C=2 k_0 k_2 -3 {k_1}^2/4+k_1 {k_2}^2/2 ,
\label{3e34}\\
\text{II)}\ &\quad k_3=2/3, \quad k_1 =3 {k_2}^2, \quad k_2, k_0:\text{arbitrary},
 \quad c_7=6 k_0-113 {k_2}^3,  
\nonumber\\
&\qquad 
C=-18 k_0 k_2 -129 {k_2}^4/4  ,
\label{3e35} \\
\text{III)}\ &\quad k_3=1/3, \quad k_2=0, \quad k_1 =0, \quad 
k_0:\text{arbitrary}, \quad c_7=33 k_0, \quad C=0 .
\label{3e36} 
\end{align}
We take the most general solution i.e.\ I) case, 
which is the same differential equation of the elliptic curve as 
that of the third order KdV equation Eq.(\ref{3e11}) 
and $c_7$ is determined as 
$c_7=-2 k_0-2k_1 k_2+{k_2}^3$.
Elliptic 1-soliton solution is given by 
\begin{equation}
u(x,t_7)=u(X^{(7)})=2 \wp(X^{(7)})-\frac{k_2}{6}  , 
\label{3e37}
\end{equation}
with
$$
X^{(7)}=x+c_7 t_7+\delta, \quad c_7=-2 k_0-2k_1 k_2+{k_2}^3  .
$$
%

\subsection{Elliptic solution for the ninth order KdV equation}

The ninth order KdV equation is given by~\cite{Shen}, 
\begin{eqnarray} 
&&u_{t_{9}}-\bigr(u_{8x}-18u u_{6x} -54  u_x u_{5x}-114 u_{2x} u_{4x}
-69{u_{3x}}^2 
+126 u^2 u_{4x}+504 u u_{x} u_{3x} \nonumber\\
&&+462 {u_x}^2 u_{2x}+378 u {u_{2x}}^2
-630 u^2 {u_x}^2-420 u^3 u_{2x}+126 u^5\bigl)_x=0 .
\label{3e39} 
\end{eqnarray}
Assuming that $x$ and $t_9$ come in the combination
of $X=x+c_9 t_9+\delta$, we have 
\begin{eqnarray} 
&&c_9u-\bigr(u_{8x}-18u u_{6x} -54  u_x u_{5x}-114 u_{2x} u_{4x}
-69{u_{3x}}^2 
+126 u^2 u_{4x}+504 u u_{x} u_{3x} \nonumber\\
&&+462 {u_x}^2 u_{2x}+378 u {u_{2x}}^2
-630 u^2 {u_x}^2-420 u^3 u_{2x}+126 u^5\bigl)+C=0 .\label{3e40} 
\end{eqnarray}
Substituting Eq.(\ref{3e1}), $\cdots$, and  Eq.(\ref{3e8})
into Eq.(\ref{3e40}) and comparing coefficients of $u^5$, $u^4$, $u^3$, 
$u^2$, $u^1$, and $u^0$,
 we have 6 conditions for 6 constants 
$k_3$, $k_2$, $k_1$, $k_0$, $c_9$, and $C$ in the following form
\begin{subequations} 
\begin{align}
\text{i)}\ &\quad (k_3-2)(3k_3-2)(3k_3-1)(5k_3-1)=0,
\label{3e41}\\ 
\text{ii)}\ &\quad k_2 (k_3-2)(3k_3-2)(3k_3-1)=0,
\label{3e42}\\
\text{iii)}\ &\quad k_1(k_3-2)(3k_3-2)(9k_3-4)
+7 {k_2}^2(k_3-2)(3k_3-2)=0 ,
\label{3e43}\\
\text{iv)}\ &\quad 3k_0(k_3-2)(225{k_3}^2-252k_3+70) \nonumber\\
&\quad +k_2(k_3-2)(720k_1k_3-546k_1 +85{k_2}^2)=0 ,
\label{3e44}\\
\text{v)}\ &\quad 
c_9=(675 {k_3}^2 -1836 k_3 +966) k_0 k_2
+(378 {k_3}^2 -1080 k_3 +651) {k_1}^2/2
\nonumber\\
&\quad 
+ (243 k_3 - 492) k_1 {k_2}^2/2 + {k_2}^4 ,
\label{3e45}\\
\text{vi)}\ &\quad 
C = (297 {k_3}^2 -828 k_3 +462) k_0 k_1/2
+ (63 k_3-123) k_0 {k_2}^2 
\nonumber\\
&\quad 
+(27 k_3 -57) {k_1}^2 k_2/2 +k_1 {k_2}^3/2 .
\label{3e46}  
\end{align}
\end{subequations}

\vspace*{-5mm}
\noindent
Then we obtain 4 solutions 
\begin{align}
\text{I)}\ &\quad 
k_3=2, \quad k_2, k_1, k_0:\text{arbitrary},  \quad  
c_9=-6 k_0 k_2 +3 {k_1}^2/2-3k_1 {k_2}^2+{k_2}^4, 
\nonumber\\
&\qquad C=-3 k_0 k_1 +3 k_0 {k_2}^2-3 {k_1}^2 k_2/2+k_1 {k_2}^3/2 ,
\label{3e47}\\
\text{II)}\ &\quad 
k_3=2/3, \quad k_0=(66 k_1 k_2 -85 {k_2}^3)/6, \quad  
k_2, k_1:\text{arbitrary}, 
\nonumber 
\\
&\qquad 
c_9=(99 {k_1}^2 +594 k_1 {k_2}^2-1188 {k_2}^4)/2, 
\quad   
C=(423 {k_1}^2 k_2 -2376 k_1 {k_2}^3 +2295 {k_2}^5)/2 ,
\label{3e48} \\
\text{III)}\ &\quad 
k_3=1/3, \quad k_1=7 {k_2}^2, \quad k_0=187{k_2}^3/3, \quad  
k_2:\text{arbitrary}, \quad  
c_9=33462 {k_2}^4, 
\nonumber 
\\
&\qquad C=40248 {k_2}^5,
\label{3e49}\\
\text{IV)}\ &\quad 
k_3=1/5, \quad k_2=0, \quad k_1 =0, \quad k_0=0, \quad c_9=0, \quad C=0 .
\label{3e50} 
\end{align}
We take the most general solution i.e.\  I) case, 
which gives the same differential equation of the elliptic 
curve as that of the third order KdV equation Eq.(\ref{3e11}), 
and $c_9$ is determined as 
$c_9=-6 k_0 k_2 +3 {k_1}^2/2-3k_1 {k_2}^2+{k_2}^4$.
Elliptic 1-soliton solution is given by
\begin{equation}
u(x,t_9)=u(X^{(9)})=2 \wp(X^{(9)})-\frac{k_2}{6} ,
\label{3e51}
\end{equation}
with
$$
X^{(9)}=x+c_9 t_9+\delta, \quad 
c_9=-6 k_0 k_2 +3 {k_1}^2/2-3k_1 {k_2}^2+{k_2}^4 .
\label{3e52}
$$

In this way, even for higher order KdV equations, the main 
structure of the elliptic solution, which is 
expressed by $X^{(2n+1)}$, takes the same functional form 
except the time dependence, that is, $c_{2n+1}$ in 
$X^{(2n+1)}=x+c_{2n+1} t_{2n+1} +\delta$.
Compared with the trigonometric/hyperbolic case,  
$c_{2n+1}$ becomes complicated for elliptic solutions 
of higher order KdV equations.

In the general $(2n+1)$-th order KdV equation, by dimensional 
analysis $[u_{2nx}]=[u^{n+1}]=M^{2n+2}$, 
integrated differential equation gives the $(n+1)$-th 
order polynomial of $u$. Then the number of the 
conditions is $n+2$, while the number of constants is 6. 
So, $n\ge5$  
becomes the overdetermined case, but we expect the 
existence of the differential 
equation of the elliptic curve 
for more than eleventh order KdV equation owing to the nice SO(2,1) Lie group symmetry. Although the existence of such elliptic curve is a priori not guaranteed,
we will show later that the elliptic solutions really exist for 
all higher order KdV equations.

\section{B\"{a}cklund transformation for the differential 
equation of the elliptic curve} 
\setcounter{equation}{0}

Here we will show that the B\"{a}cklund transformation connects 
one solution to another solution of the same differential equation of 
the Weierstrass type elliptic curve.
The Lie group structure of KdV equation is given by  
GL(2,$\mathbb{R}$) $\cong$ SO(2,1) and the B\"{a}cklund transformation 
can be considered as the self gauge 
transformation of this Lie group.  
We consider two elliptic solutions for the KdV equation, that is, 
two solutions $u'(x, t_3)$ and $u(x, t_3)$ for $u'_{t_3}-u'_{xxx}+6 u' u'_x=0$ and 
$u_{t_3}-u_{xxx}+6 u u_x=0$. We put the time dependence in the forms;   
$X'=x+c'_3 t_3+\delta'$ for $u'(x, t_3)$ and that of $X=x+c_3 t_3+\delta$ 
for $u(x,t_3)$. In order to connect two solutions by the B\"{a}cklund 
transformation and to construct $N$-soliton solutions, 
$c'_3$ and $c_3$ must take the same common value. 
By integrating twice, we have the same differential equation of the elliptic 
curve
\begin{align}
{u'_{x}}^2&=2 u'^3+k_2 u'^2+k_1 u'+k_0 ,
\label{4e1}\\
{u_{x}}^2&=2u^{3}\hphantom{'}+k_2 u^2\hphantom{'}+k_1 u\hphantom{'}+k_0, 
\label{4e2}
\end{align}
with same coefficients $k_2$, $k_1$, and $k_0$, where we 
take $c_3=c'_3=k_2$.
By taking a constant shift of $u \rightarrow u-k_2/6$, we consider 
the same two differential equations of the Weierstrass type elliptic curve
\begin{align}
{u'_{x}}^2&=2 u'^3-2g_2 u' -4g_3 ,
\label{4e3}\\
{u_{x}}^2&=2u^3\hphantom{'}-2g_2 u\hphantom{'} -4g_3, 
\label{4e4}
\end{align}
where $g_2$ and $g_3$ are given by Eqs.(\ref{3e13}) and (\ref{3e14}). 
It should be mentioned that this differential equation of the
Weierstrass 
type elliptic curve has not only the solution $u(x)=2\wp(x)$ but also
$N$-soliton solutions~\cite{Hayashi2}.

Here we will show that we can connect two 
solutions of Eqs.(\ref{4e3}) and (\ref{4e4})
by the following B\"{a}cklund transformation
\begin{equation}
z'_x+z_x=-\frac{a^2}{2}+\frac{(z'-z)^2}{2}  ,
\label{4e5}
\end{equation} 
where $u=z_x$ and $u' =z'_x$. 
We introduce $U=u'+u=z'_x+z_x$ and $V=z'-z$, 
which gives $V_x=z'_x-z_x=u'-u$.
Then we have $u'=(U+V_x)/2$ and $u=(U-V_x)/2$. 
Eqs.(\ref{4e3}) and (\ref{4e4}) are given by
\begin{align}
(U_x+V_{xx})^2&=(U+V_x)^3-4g_2(U+V_x)-16g_3 ,
\label{4e6}\\
(U_x-V_{xx})^2&=(U-V_x)^3-4g_2(U-V_x)-16g_3 .
\label{4e7}
\end{align}
The B\"{a}cklund transformation (\ref{4e5}) is given by
\begin{equation}
U=\frac{V^2}{2}-\frac{a^2}{2}  ,
\label{4e8}
\end{equation}
which gives $U_x=V V_x$.

First, by taking Eq.(\ref{4e6})$-$Eq.(\ref{4e7}), we have
\begin{equation}
U_xV_{xx}=\frac{1}{2}\left(3U^2V_x+{V_x}^3\right)-2 g_2 V_x  ,
\label{4e9}
\end{equation}
which reads the form 
\begin{equation}
V V_{xx}=\frac{3}{8}\left(V^2-a^2\right)^2+\frac{1}{2} {V_x}^2 -2 g_2
= \frac{1}{2}{V_x}^2+\frac{3}{8}V^4-\frac{3}{4}a^2 V^2 +\frac{3}{8}a^4-2
g_2, 
\label{4e10}
\end{equation}
through the relation (\ref{4e8}). 
By dimensional analysis, we have 
\begin{equation}
V_x^2=m_4 V^4+m_3 V^3+ m_2 V^2+m_1 V+m_0 ,
\label{4e11}
\end{equation}
where $m_i (i=0, 1, \cdots, 4)$ are constants.
By differentiating this relation, we have
\begin{equation}
V_{xx}=2m_4 V^3+\frac{3}{2} m_3 V^2+ m_2 V+\frac{1}{2} m_1 .
\label{4e12}
\end{equation}
Substituting this relation into Eq.(\ref{4e10}), we have
\begin{equation}
2 m_4 V^4+\frac{3}{2} m_3 V^3+ m_2 V^2+\frac{1}{2} m_1 V
=\frac{1}{2} {V_x}^2+\frac{3}{8} V^4-\frac{3}{4} a^2 V^2 
+\frac{3}{8}a^4-2 g_2 ,   \label{4e13}
\end{equation}
which gives
\begin{align}
{V_x}^2&=\left(4m_4-\frac{3}{4}\right) V^4+3 m_3 V^3+ 
\left(2 m_2+\frac{3}{2}a^2\right) V^2+m_1V
-\frac{3}{4}a^4+4 g_2 \nonumber\\
&=m_4 V^4+m_3 V^3+ m_2 V^2+m_1 V+m_0  .
\label{4e14}
\end{align}
Comparing coefficients of the power of $V$, 
we have $m_4=1/4$, $m_3=0$,
$m_2=-3a^2/2$, $m_1=(\text{undetermined})$, 
$m_0=-3a^{4}/4+4g_2$, 
which gives
\begin{align}
{V_x}^2&=\frac{1}{4} V^4-\frac{3}{2} a^2 V^2+m_1 V
-\frac{3}{4} a^{4}+4 g_2 ,
\label{4e15}\\
V_{xx}&=\frac{1}{2} V^3-\frac{3}{2} a^2 V+\frac{1}{2} m_1.
\label{4e16}
\end{align}

Second, by taking Eq.(\ref{4e6})$+$Eq.(\ref{4e7}), we have
\begin{equation}
{U_x}^2+{V_{xx}}^2=U^3+3 U {V_x}^2-4 g_2 U-16 g_3 .
\label{4e17}
\end{equation}
Using Eq.(\ref{4e8}), we have 
\begin{eqnarray}
V^2 {V_x}^2+{V_{xx}}^2=\left(\frac{V^2}{2}-\frac{a^2}{2}\right)^{3}
+3\left(\frac{V^2}{2}-\frac{a^2}{2}\right){V_x}^2 
-4 g_2\left(\frac{V^2}{2} - \frac{a^2}{2}\right)-16g_3.
\label{4e18}
\end{eqnarray} 
Substituting ${V_x}^2$ and $V_{xx}$ into Eq.(\ref{4e18}) and 
by using Eq.(\ref{4e15}) and Eq.(\ref{4e16}), we have the 
condition ${m_1}^2=4 a^6-16a^2g_2-64 g_3$. Then the undetermined 
coefficient $m_1$ is determined, and we have the differential equation of 
the Jacobi type elliptic curve for $V=z'-z$
\begin{equation}
{V_x}^2=\frac{1}{4}V^4-\frac{3}{2} a^2 V^2
\pm\sqrt{4 a^6-16a^2g_2-64 g_3}\,V
-\frac{3}{4}a^{4}+4g_2  .
\label{4e19}
\end{equation}

In this way, the set of equations $\Big\{{\rm Eq}.(\ref{4e3}), \ 
{\rm Eq}.(\ref{4e5})\Big\}$ is equivalent to 
the set of those $\Big\{{\rm Eq}.(\ref{4e4}), \ {\rm Eq}.(\ref{4e5})\Big\}$.  
This means that the B\"{a}cklund transformation (\ref{4e5}) connects 
one soliton solution $u$ to another soliton solution $u'$ 
for the same differential equation Eq.(\ref{4e3}) and Eq.(\ref{4e4}) 
of the Weierstrass type elliptic curve. In order to construct $N$-soliton 
solutions of the $(2n+1)$-th order KdV equation by the B\"{a}cklund
 transformation, the time dependence for each 1-soliton solution, 
${c_{2n+1}}_i \, (i=1, 2, \cdots, N)$, must be the same common value, 
then $x$ and $t_{2n+1}$ come in the combination
$X^{(2n+1)}_i=x+c_{2n+1} t_{2n+1}+\delta_i$.

In our previous work~\cite{Hayashi2}, by using the explicit soliton solution 
given by $\wp$-function and $\zeta$-function, we connect 
one soliton solution to another soliton solution by the B\"{a}cklund transformation.
Here we have shown that B\"{a}cklund 
transformation connects one soliton solution to another soliton solution
of the same differential equation of the Weierstrass type elliptic 
curve without using the explicit expression of the solution.

\section{Special hyperelliptic solutions for higher order KdV equations} 
\setcounter{equation}{0}

By using the method of commutative ordinary operators~\cite{Burchnall,Date2}, we
can formulate higher order KdV equations into the Jacobi's inversion problem.
By solving the general Jacobi's inversion problem, we can find solutions for  
higher order KdV equations~\cite{McKean,Date2,Its,Dubrobin,Krichever}.
Here we consider the fifth order KdV equation in order to explain how to 
solve the Jacobi's inversion problem. 
Integrated fifth order KdV equation is given by
\begin{eqnarray}
u_{4x}-10u u_{2x} -5{u_{x}}^2+10u^3=c_5 u+C.  
\label{5e1} 
\end{eqnarray}
According to the Tanaka-Date's nice paper~\cite{Date2}, this fifth order KdV
equation is reformulated in the following form. 
We introduce auxiliary fields $\mu_1(x), \mu_2(x)$, 
\begin{align}
u(x)&=2(\mu_1(x)+\mu_2(x)) ,
\label{5e2}\\
\mu_1(x)_x&=\frac{\pm 2\sqrt{f_5(\mu_1(x))}}{\mu_1(x)-\mu_2(x)} ,
\label{5e3}\\
\mu_2(x)_x&=\frac{\pm 2\sqrt{f_5(\mu_2(x))}}{\mu_2(x)-\mu_1(x)} ,
\label{5e4}\\
f_5(\lambda)&=\lambda^5+\alpha_3\lambda^3+\alpha_2\lambda^2
+\alpha_1\lambda +\alpha_0 ,
\label{5e5} 
\end{align}
where $\alpha_3$, $\alpha_2$, $\alpha_1$, and $\alpha_0$ are constants. 
Surprisingly, this $u(x)$ satisfies 
\begin{equation}
u_{4x}-10u u_{2x} -5{u_{x}}^2+10u^3=-8\alpha_3 u+16\alpha_2   .
\label{5e6} 
\end{equation}
which determines $c_5=-8 \alpha_3, C=16 \alpha_2$.
Then if we can find the solution $\mu_1(x), \mu_2(x)$, we can construct 
the solution $u(x,t)$ of the fifth order KdV equation by 
$u(x,t)=u(X^{(5)})=2\left(\mu_1(X^{(5)})+\mu_2(X^{(5)})\right)$ where
$X^{(5)}=x+c_5 t_5+\delta$.

Eqs.(\ref{5e3}) and (\ref{5e4}) can be written in the form of the 
genus two Jacobi's inversion problem~\cite{Shigemoto3}
\begin{align}
\frac{{\rm d}\mu_1(x)}{\sqrt{f_5(\mu_1(x))}}
+\frac{{\rm d}\mu_2(x)}{\sqrt{f_5(\mu_2(x))}}&=0  ,
\label{5e7}\\
\frac{\mu_1(x)\, {\rm d}\mu_1(x)}{\sqrt{f_5(\mu_1(x))}}
+\frac{\mu_2(x)\, {\rm d}\mu_2(x)}{\sqrt{f_5(\mu_2(x))}}
&=\pm 2\, {\rm d}x    .
\label{5e8}
\end{align}
The solution of the Jacobi's inversion problem is that the symmetric combination of 
$\mu_1(x)$ and $\mu_2(x)$, that is , $\mu_1(x)+\mu_2(x)(=u(x)/2)$ 
and $\mu_1(x) \mu_2(x)$ are given 
by the ratio of the genus two hyperelliptic theta function. 
However, the above 
Jacobi's inversion problem
is special as the right-hand side of Eq.(\ref{5e7}) is zero.
Then the genus two hyperelliptic theta function takes in the following special 
1-variable form 
$\vartheta(\pm 2x+d_1, d_2)$ where $d_1, d_2$ are constants, that is,  
the second argument 
becomes constant. Then the ratio of such
special genus two hyperelliptic theta function is the function 
of 1-variable $x$, 
which becomes proportional to the 1-variable 
function $u(x)=2(\mu_1(x)+\mu_2(x))$. 
The general genus two hyperelliptic theta function is given by 
\begin{equation}
\vartheta(u,v;\tau_1,\tau_2,\tau_{12})=\sum_{m,n \in \mathbb{Z} }
\exp \Big[ i \pi(\tau_1 m^2+\tau_2 n^2+2 \tau_{12} m n)+2 i \pi (m u+n v) \Big] .
\label{5e9}
\end{equation}
Then $F(x,t)=\vartheta(x, d_2; t, \tau_2, \tau_{12})$ satisfies the diffusion equation 
$\partial_t F(x, t)=-i \partial_x^2 F(x, t)/4 \pi$.
Further, $F(x, t)$ has the trivial periodicity $F(x+1, t)=F(x, t)$. 
It is shown in the Mumford's nice textbook~\cite{Mumford} that 
if $F(x,t)$ satisfies 
i) periodicity $F(x+1, t)=F(x, t)$, \, ii) diffusion equation  
$\partial_t F(x, t)=-i \partial_x^2 F(x, t)/4 \pi$, $F(x, t)$ becomes the genus one  
elliptic theta function of 1-variable $x$. 
By solving the Jacobi's inversion problem, the 
solution $u(x, t_5)=u(X^{(5)})=u(x+c_5 t_5+\delta)$ of 
the fifth order KdV equation is given by the ratio of the special 
1-variable hyperelliptic 
theta function, which gives the elliptic solution.
For the $(2n+1)$-th order KdV equation, the solution of the Jacobi's 
inversion problem gives 
$u(x, t_{2n+1})=u(X^{(2n+1)})$ as the ratio of the special 1-variable genus 
$n$ hyperelliptic theta 
function of the form 
$\vartheta(\pm 2x+d_1, d_2, \cdots, d_n)$, which also becomes 
the genus one elliptic theta function.

For higher order KdV equations, it is shown that solutions
are expressed with above 
special 1-variable hyperelliptic theta functions, which becomes  
elliptic theta functions.
Then we can conclude that all higher order KdV equations always have  
elliptic solutions, though we have explicitly constructed elliptic solutions 
only up to the ninth order KdV equation.

\section{Summary and Discussions} 
\setcounter{equation}{0}

We have studied to construct $N$-soliton solution for the 
Lax type higher order KdV equations 
by using the GL(2,$\mathbb{R}$) $\cong$ SO(2,1) Lie group structure.
The main structure of $N$-soliton solutions, expressed with 
$X_i=\alpha_i x+\beta_i t +\delta_i, (i=1,2, \cdots, N)$ is the same even 
for higher order KdV equations. The difference of $N$-soliton solutions 
in various higher order KdV equations 
is the time dependence, that is, coefficients $\beta_i$.

In trigonometric/hyperbolic solutions, by taking the Lie algebra limit,
we can easily determine the time dependence.
For the $(2n+1)$-th order KdV equation, we can 
obtain $N$-soliton solutions from those of the original KdV equation
by just the replacement 
$X^{(3)}_i=a_i x+a_i^3 t_3+\delta_i $ $\rightarrow$
$X^{(2n+1)}_i=a_i x+a^{2n+1}_i t_{2n+1}+\delta_i$, $(i=1,2, \cdots, N)$.

For elliptic solutions, up to the ninth order KdV equation, we have  
obtained $N$-soliton solutions from those of the original KdV equation
by just the replacement 
${X^{(3)}}_i=x+c_3 t_3+\delta_i $ $\rightarrow$
${X^{(2n+1)}}_i=x+c_{2n+1} t_{2n+1}+\delta_i$, $(i=1, 2, 3, 4)$ 
where $c_{2n+1}$ are given by
$c_3=k_2$, $c_5=-k_1+{k_2}^2$, $c_7=-2 k_0-2k_1 k_2+{k_2}^3$, and  
$c_9=-6 k_0 k_2 +3 {k_1}^2/2-3k_1 {k_2}^2+{k_2}^4$ by using coefficients
of differential equation of the Weierstrass type elliptic curve 
${u_x}^2=2 u^3+k_2 u^2+k_1 u+k_0$.
 
For general higher order KdV equations, 
equations becomes quite complicated,
and it became difficult to use our method to show that 
elliptic solutions always exist. But we can show that the 
elliptic solution for all higher order 
KdV equation always exists by the following 
two different ways.

First way is to use the GL(2,$\mathbb{R}$)$\cong$ SO(2,1)
Lie group structure.
For all higher order KdV equations, 
we have the same GL(2,$\mathbb{R}$)$\cong$ SO(2,1) 
Lie group structure and the same B\"{a}cklund transformation,
which means that the main structure expressed with 
the variable ${X^{(2n+1)}}=x+c_{2n+1} t_{2n+1}+\delta$ is the same and 
difference is only the time dependence $c_{2n+1} $.
Then, as the elliptic solution of the third order KdV equation exist
with ${X^{(3)}}$ variable, the existence of the elliptic solution
of all higher order KdV equation with ${X^{(2n+1)}}$ is guaranteed.

Second way is to formulate in the Jacobi's inversion problem.
For the general $(2n+1)$-th order KdV equation, it can be formulated in the 
Jacobi's inversion problem~\cite{Burchnall,Date2}, and it is known 
that there exist 
solutions expressed with the special 1-variable hyperelliptic theta 
function of the form 
$\vartheta(\pm 2x+d_1, d_2, \cdots, n)$~\cite{Date2,McKean,Its,Dubrobin,Krichever}, 
which is shown to be the elliptic theta function according to the 
Mumford's argument~\cite{Mumford}. 
We can say in another way. As the soliton solution $u(x, t)=u(X)$, 
$(X=\alpha x+\beta t_{2n+1}+\delta)$, which is expressed 
as the ratio of special 1-variable hyperelliptic theta functions, as it has 
the trivial periodicity $X \rightarrow X+1$, $u(X)$ must be the 
trigonometric/hyperbolic or the elliptic function. 
Then it becomes the elliptic function according to the 
Mumford's argument.

By using these two different ways, 
we can conclude that we always have the elliptic solutions 
for the general higher order KdV equations.

Further, without using the explicit form of the solution expressed with the 
$\wp$ function, we have shown that the KdV type B\"{a}cklund 
transformation connects one solution to another solution of the same 
differential equation of the Weierstrass type elliptic curve.

.

\end{document}